\documentclass[aps,prd,preprint]{revtex4}
\usepackage{amsmath,amssymb,graphics,epsfig,subfigure}
\usepackage{color}
\usepackage[colorlinks,linkcolor=red,anchorcolor=red,citecolor=green]{hyperref}
\usepackage{setspace}
\begin{document}

\thispagestyle{empty}

\begin{center}

\title{Ruppeiner thermodynamic geometry for the Schwarzschild-AdS black hole}

\author{Zhen-Ming Xu$^{1,2,3,}$\footnote{E-mail: xuzhenm@nwu.edu.cn}, Bin Wu$^{1,2,3,}$\footnote{E-mail: binwu@nwu.edu.cn {\color{red}(Corresponding author)}}
        and Wen-Li Yang$^{1,2,3,4,}$\footnote{E-mail: wlyang@nwu.edu.cn}
        \vspace{5pt}\\}

\affiliation{ $^{1}$Institute of Modern Physics, Northwest University, Xi'an 710127, China\\
$^{2}$School of Physics, Northwest University, Xi'an 710127, China\\
$^{3}$Shaanxi Key Laboratory for Theoretical Physics Frontiers, Xi'an 710127, China\\
$^{4}$NSFC-SFTP Peng Huanwu Center for Fundamental Theory, Xi'an 710127, China}

\begin{abstract}
Due to the nonindependence of entropy and thermodynamic volume for spherically symmetric black holes in the AdS spacetime, when applying the Ruppeiner thermodynamic geometry theory to these black holes, we often encounter an unavoidable problem of the singularity about the line element of thermodynamic geometry. In this paper, we propose a basic and natural scheme for dealing with the thermodynamic geometry of spherically symmetric AdS black holes. We point out that enthalpy, not internal energy, is the fundamental thermodynamic characteristic function for the Ruppeiner thermodynamic geometry. Based on this fact, we give the specific forms of the line element of thermodynamic geometry for Schwarzschild AdS (SAdS) black hole in different phase spaces and the results show that the thermodynamic curvatures obtained in different phase spaces are equivalent. It is shown that the thermodynamic curvature is negative which may be related to the information of attractive interaction between black hole molecules for the SAdS black hole. Meanwhile we also give an approximate expression of the thermodynamic curvature of the Schwarzschild black hole which shows that the black hole may be dominated by repulsion on low temperature region and by attraction on high temperature region phenomenologically or qualitatively.
\end{abstract}

\maketitle
\end{center}

\section{Motivation}
``If you can heat it, it has microscopic structure", the view of Boltzmann provides a good basis to determine whether the system has microstructure or not. With the foundation works of Hawking and Bekenstein that the black hole has temperature and entropy on an event horizon \cite{Hawking1975,Bekenstein1973}, there is no doubt that a black hole has microstructure. This issue plays a decisive role in the study of black hole physics and even gravitation theory. The development of black hole thermodynamics contributes to exploring the microscopic state of black hole more intuitively and conveniently \cite{Wald2001,Hawking1983,Padmanabhan2010,Carlip2014,Chamblin1999}. Especially in recent years, the proposition of the thermodynamic geometry of black hole \cite{Ruppeiner1995,Ruppeiner2008,Ruppeiner2010,Ruppeiner2014,Cai1999,Zhang2015,Zhang2015b,Liu2010,Bhattacharya2019} and the hypothesis of black hole molecule \cite{Wei2015} have promoted the study of the microstructure of black hole completely from the thermodynamic point of view.

What's more interesting is that the introduction of the extended phase space with a pair of new conjugate quantities of the thermodynamic pressure $P$ and thermodynamic volume $V$ \cite{Kastor2009}. Black holes exhibit abundant phase transition behavior and microstructure in the extended phase space \cite{Dolan2011a,Dolan2011b,Kubiznak2012,Kubiznak2017}. The key of introducing extended phase space is to interpret the cosmological constant $\Lambda$ as the thermodynamic pressure $P$ with
\begin{equation}\label{pressure}
P=-\frac{\Lambda}{8\pi}=\frac{3}{8 \pi l^2 },
\end{equation}
where $l$ represents the curvature radius of the AdS spacetime. Then the black hole mass $M$ can be identified with the enthalpy, rather than the internal energy \cite{Kastor2009}.

For spherically symmetric black holes in the AdS spacetime, like SAdS black hole, Reissner-Nordstr\"{o}m AdS (RN-AdS) black hole, Gauss-Bonnet AdS (GB-AdS) black hole and etc., with the introduction of extended phase space $\{P,V\}$, the biggest difference between the thermodynamic properties of these black holes and those of ordinary thermodynamic systems is that these black holes thermodynamic systems have a zero heat capacity at constant volume, i.e., $C_{_V}:=T(\partial S/\partial T)_{_V}=0$, where $S$ is the entropy and $T$ is Hawking temperature of black hole. The reason is that the entropy $S$ and thermodynamic volume $V$ of these black holes are not independent (both of them are just functions of the horizon radius). This is also the unique property of black hole thermodynamics. When we apply the Ruppeiner thermodynamic geometry theory to the kind black hole system, we should be careful, because there are some subtle differences in the thermodynamic geometry of these black holes compared with those without AdS background. Due to a vanishing heat capacity at constant volume or non-independence of entropy and thermodynamic volume, this renders the line element of Ruppeiner thermodynamic geometry singular, which brings about a divergent thermodynamic curvature. Consequently some micro information of the associated black hole is not revealed from the thermodynamic geometry. One of the feasible solutions is the introduction of normalized thermodynamic curvature proposed by the authors \cite{Wei2019a,Wei2019b,Wei2019c} by treating the heat capacity at constant volume as a constant very close to zero. Thus, the microscopic behaviors of RN-AdS and GB-AdS black holes have been analyzed in detail with the help of the normalized thermodynamic curvature.

Now in present paper, we propose another feasible, more direct and natural solution to the problem of singularity about the line element of Ruppeiner thermodynamic geometry caused by non-independence of entropy and thermodynamic volume. This scheme is probably the most fundamental for the thermodynamic geometric analysis of spherically symmetric black holes in the AdS spacetime. Our starting point is that the thermodynamic differential relation of internal energy $dU=TdS-PdV$ will no longer hold because entropy and thermodynamic volume are not independent and the most basic thermodynamic differential relation for such spherically symmetric black holes in the AdS spacetime is about that of enthalpy $M$,
\begin{eqnarray}\label{law}
dM=TdS+VdP+\text{other works}.
\end{eqnarray}
Based on the fundamental relationship Eq.~(\ref{law}), we give the general form of the line element of Ruppeiner thermodynamic geometry for SAdS black hole and the specific forms in phase space $\{S,P\}$ and $\{T,V\}$. The results show that thermodynamic curvatures obtained in the two phase spaces are equivalent, and the thermodynamic curvature is always negative which may be related to the information of attractive interaction between black hole molecules for the SAdS black hole. Meanwhile we also give an approximate expression of the thermodynamic curvature of the Schwarzschild black hole, which shows that the black hole is dominated by repulsion on low temperature region and by attraction on high temperature region phenomenologically or qualitatively.

\section{Ruppeiner thermodynamic geometry}
The Ruppeiner thermodynamic geometry, which based on the fluctuation theory of equilibrium thermodynamics, is established on the language of Riemannian geometry \cite{Ruppeiner1995}. Now it is dealt with as a new attempt to extract the microscopic interaction information from the axioms of thermodynamics \cite{Ruppeiner2008,Ruppeiner2010,Ruppeiner2014}. Its line element can be written as in terms of entropy representation
\begin{equation}\label{line}
\Delta l^2=-\frac{\partial^2 S}{\partial X^{\mu}\partial X^{\nu}}\Delta X^{\mu}\Delta X^{\nu},
\end{equation}
where $X^{\mu}$ represents some independent thermodynamic quantities. In ordinary thermodynamics, for some better understood statistical mechanical models, it is an empirical observation that negative (positive) thermodynamic curvature is associated with attractive (repulsive) microscopic interactions \cite{Ruppeiner2010}. For black hole systems, it is observed in \cite{Ruppeiner2008,Ruppeiner2014,Wei2015,Wei2019a,Wei2019b,Wei2019c} that there should be similar results with ordinary thermodynamic systems. In \cite{Dolan2015}, the author Dolan analyzed the thermodynamic geometry in internal energy and entropy representations under the extended phase space framework and pointed out for the first time that it is not at all clear whether or not this interpretation (attractive microscopic forces tend to give negative Ruppeiner curvature while repulsive forces give positive curvature) accounts for the changing sign of thermodynamic curvature for black holes. Because there does not exist hitherto an underlying theory of quantum gravity, the exploration on the microscopic structure of black holes is bound to some speculative assumptions. Based on the well-established black hole thermodynamics, as an analogy analysis and a primary description of the micro behavior of black holes, it can be said that the Ruppeiner thermodynamic geometry phenomenologically or qualitatively provides the information about interactions of black holes, like the reports in \cite{Ruppeiner2008,Ruppeiner2014,Wei2015,Wei2019a,Wei2019b,Wei2019c}, i.e., negative (positive) curvature may be related to the information of attractive (repulsive) interaction between black hole molecules. Meanwhile, the value of the thermodynamic curvature measures the strength of the interactions in some sense \cite{Miao2018}. In \cite{Ariza2014}, authors claim that thermodynamic flatness is not a sufficient condition to establish the absence of interactions in the underlying microscopic model of a thermodynamic system and propose an alternative energy representation for Kerr-Newman black holes which results that thermodynamic curvature diverges only at absolute zero. They pointed out that a criterion for the choice of an appropriate energy representation in an arbitrary thermodynamic system (or equivalently, a coordinate-free definition of Ruppeiner metric) is still missing. Here in present paper, as an attempt, we propose the enthalpy and Helmholtz free energy representations to deal with the static spherically symmetric black holes in the AdS spacetime and the results show that thermodynamic curvatures obtained in these two representations are equivalent.

Now we consider the four-dimensional SAdS black hole and its metric is \cite{Dolan2011a}
\begin{equation}
d s^2=-f(r)dt^2+\frac{d r^2}{f(r)}+r^2(d\theta^2+\sin^2 \theta d\varphi^2),
\end{equation}
and here the function $f(r)$ is
\begin{equation*}
f(r)=1-\frac{2M}{r}+\frac{r^2}{l^2}.
\end{equation*}
The basic thermodynamic quantities of the SAdS black hole are listed below in terms of the horizon radius $r_h$ which is regarded as the largest root of equation $f(r)=0$ \cite{Dolan2011a,Miao2019,Belhaj2015}
\begin{eqnarray}
\text{Temperature}&:&T=\frac{1}{4\pi r_h}+2P r_h,\label{Temperature}\\
\text{Entropy}&:&S=\pi r_h^2,\label{Entropy}\\
\text{Thermodynamic Volume}&:&V=\frac{4\pi r_h^3}{3}, \label{Volume}
\end{eqnarray}
and the heat capacity at constant pressure $C_{_P}=-2S(1+8PS)/(1-8PS)$ which first appeared in \cite{Dolan2011a}. Furthermore the most basic thermodynamic differential relation is $dM=TdS+VdP$ and we adjust this relation to get
\begin{eqnarray}
dS=\frac{1}{T}dM-\frac{V}{T}dP.
\end{eqnarray}
Now we set $X^{\mu}=(M,P)$, and then the conjugate quantities corresponding to $X^{\mu}$ are $Y_{\mu}=\partial S/\partial X^{\mu}=(1/T,-V/T)$. Hence the line element Eq.~(\ref{line}) becomes $\Delta l^2=-\Delta Y_{\mu} \Delta X^{\mu}$. After some simple mathematical derivation, we can write the line element Eq.~(\ref{line}) as a universal form for the SAdS black hole
\begin{equation}\label{uline}
\Delta l^2=\frac{1}{T}\Delta T \Delta S+\frac{1}{T}\Delta V \Delta P.
\end{equation}

(i) When the phase space is $\{S,P\}$, the line element takes the form in terms of enthalpy representation
\begin{equation}\label{linesp}
\Delta l^2=\frac{1}{C_{_P}}\Delta S^2+\frac{2}{T}\left(\frac{\partial T}{\partial P}\right)_S \Delta S \Delta P+\frac{1}{T}\left(\frac{\partial V}{\partial P}\right)_S \Delta P^2,
\end{equation}
where $C_{_P} :=T(\partial S/\partial T)_{_P}$ and we have used the Maxwell relation $(\partial T/\partial P)_{_S}=(\partial V/\partial S)_{_P}$ based on the thermodynamic differential relation $dM=TdS+VdP$. The third term in the above Eq.~(\ref{linesp}) equals zero because the non-independence of entropy and thermodynamic volume, i.e., $(\partial V/\partial P)_{_S}=0$. Nevertheless, the line element Eq.~(\ref{linesp}) is still well and has no singularity.

(ii) When the phase space is $\{T,V\}$, the line element reads as in terms of Helmholtz free energy representation
\begin{equation}\label{linetv}
\Delta l^2=\frac{C_{_V}}{T^2} \Delta T^2+\frac{2}{T}\left(\frac{\partial P}{\partial T}\right)_V \Delta T \Delta V+\frac{1}{T}\left(\frac{\partial P}{\partial V}\right)_T \Delta V^2,
\end{equation}
where we have used the Maxwell relation $(\partial S/\partial V)_{_T}=(\partial P/\partial T)_{_V}$ based on the thermodynamic differential relation $dF=-SdT-PdV$ and $F$ is Helmholtz free energy. The first term in the above Eq.~(\ref{linetv}) vanishes due to the zero heat capacity at constant volume, i.e., $C_{_V}=0$. Anyway, the line element Eq.~(\ref{linetv}) is still well.

Then we return to the SAdS black hole. With the help of Eqs.~(\ref{Temperature}),~(\ref{Entropy}) and~(\ref{Volume}), we can obtain the thermodynamic scalar curvature of  in different phase spaces.

In the phase space $\{S,P\}$, we have
\begin{equation}\label{curvaturesp}
R_{_{SP}}=-\frac{1}{S(1+8PS)}.
\end{equation}

In the phase space $\{T,V\}$, we get
\begin{equation}\label{curvaturetv}
R_{_{TV}}=-\frac{1}{3\pi T V}.
\end{equation}

Meanwhile we can directly find $R_{_{SP}}=R_{_{TV}}$. Hence for convenience, we label the two curvatures as $R$. It is clear that the curvature is negative, i.e., $R<0$, which may be related to the information of attractive interaction between black hole molecules for the SAdS black hole.

\begin{figure}
\begin{center}
\includegraphics[width=70mm]{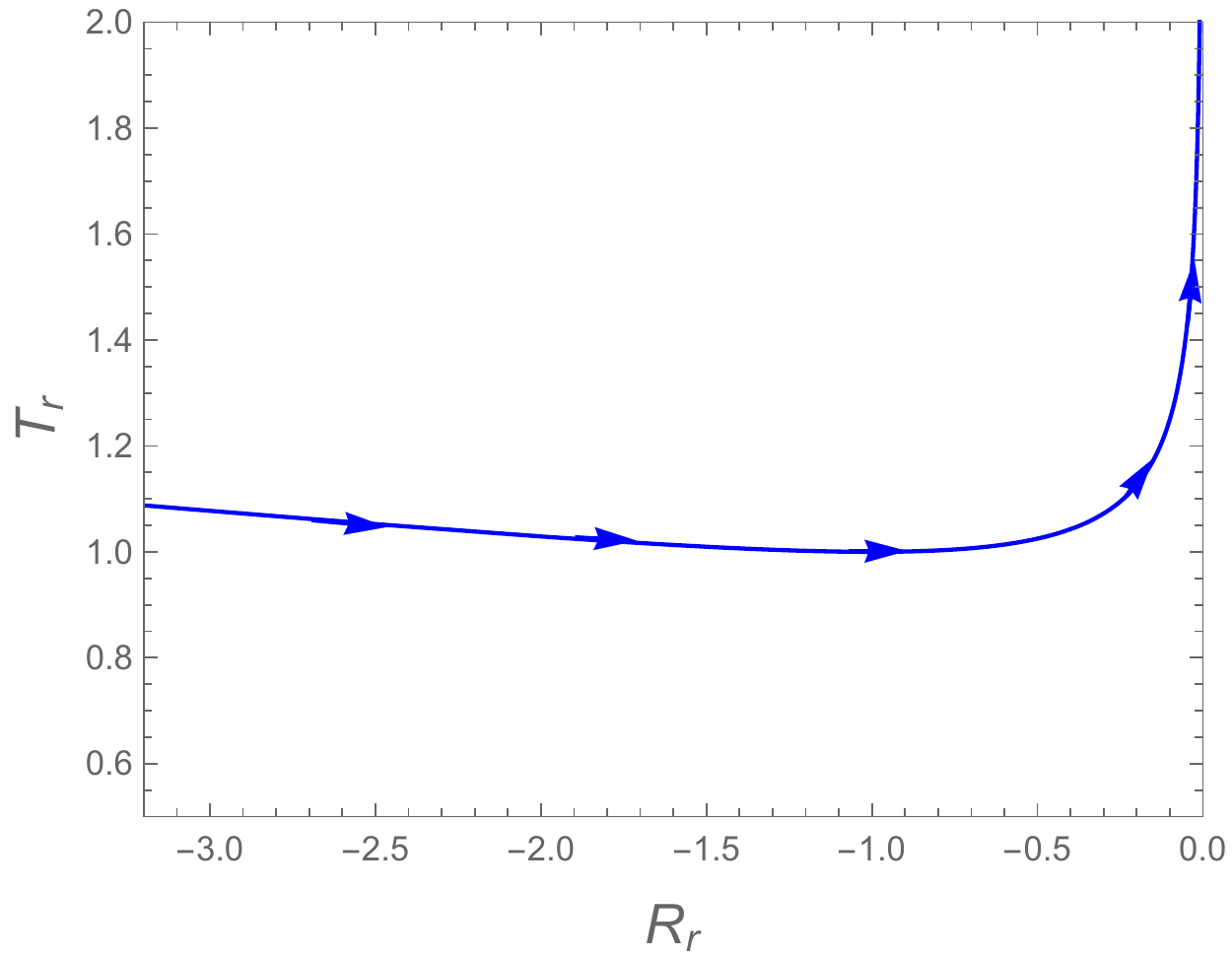}
\end{center}
\caption{The diagram of the rescaled temperature $T_r$ with respect to the rescaled thermodynamic scalar curvature $R_r$ and arrows indicate the direction in which $x$ increases for SAdS black hole.}
\label{fig1}
\end{figure}

Next we consider the case of fixed AdS spacetime, i.e., $P=\text{constant}$, and analyse the relationship between thermodynamic curvature and temperature. After a simple calculation, we have rescaled temperature $T_r$ and rescaled thermodynamic scalar curvature $R_r$
\begin{equation}\label{reducetr}
T_r=\sqrt{\frac{\pi}{2P}}T=\frac{x^{\frac12}+x^{-\frac12}}{2}, \qquad R_r=\frac{R}{4P}=-\frac{2}{x(x+1)},
\end{equation}
where $x=8PS$. Here, along the direction of increasing $x$, we show the curve of temperature $T_r$ and curvature $R_r$ in FIG. \ref{fig1}. With the decrease of absolute value of curvature, i.e., the interaction gradually weakening, the black hole temperature shows a trend of decreasing first and then increasing sharply. The increase of temperature means that the irregular free motion of black hole molecules is dominant, while the decrease of temperature means that the interaction is significant, which will inevitably suppress the temperature.

\begin{figure}
\begin{center}
\includegraphics[width=70mm]{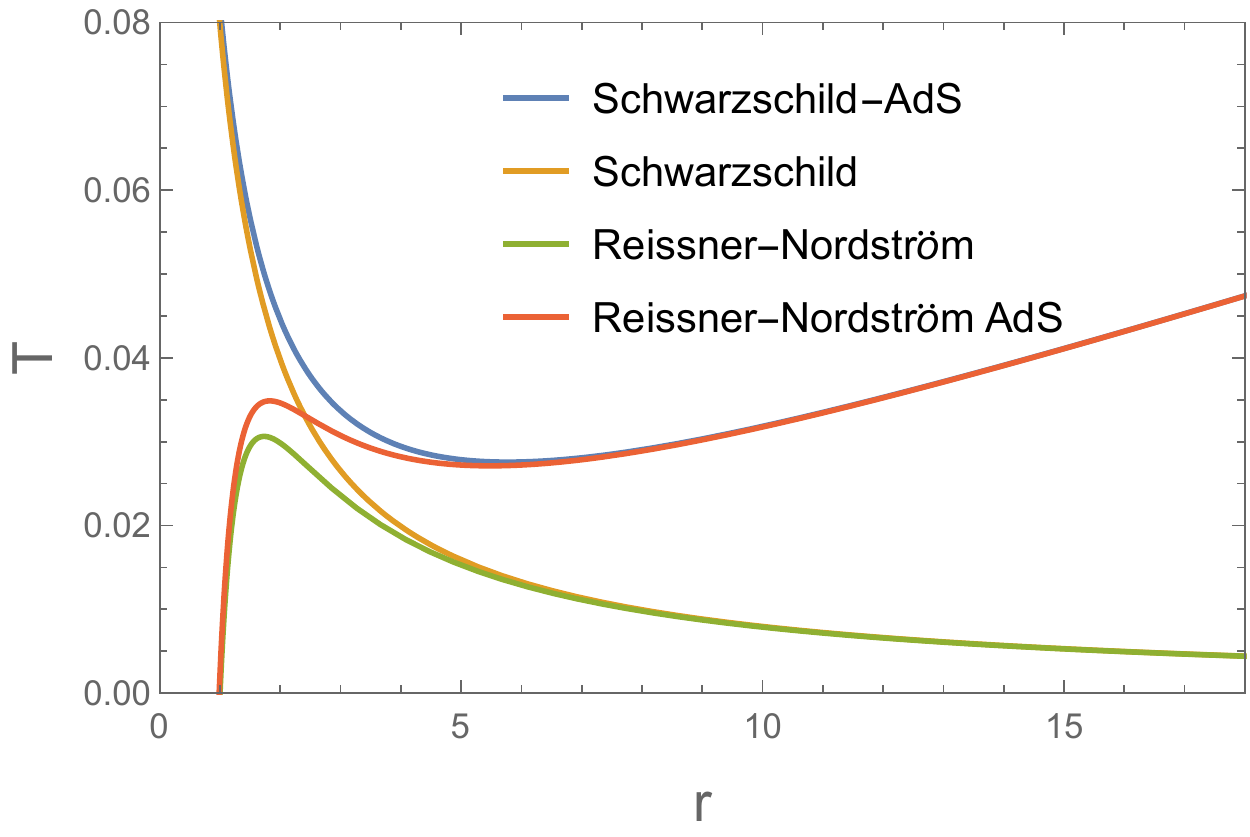}
\end{center}
\caption{The diagram of the temperature with respect to horizon radius for Schwarzschild black hole ($P=0, Q=0$), SAdS black hole ($P=3/(800\pi), Q=0$), RN black hole ($P=0, Q=1$) and RN-AdS black hole ($P=3/(800\pi), Q=1$).}
\label{fig2}
\end{figure}

Finally we turn to the thermodynamic micro-behavior of Schwarzschild black hole by means of the Ruppeiner thermodynamic geometry. Regardless of how the phase space is chosen, the line element of thermodynamic geometry of the Schwarzschild black hole is always singular, and we have to analyze its micro-behavior from the thermodynamic geometry with the help of the results of other black holes, like the SAdS black hole we are concerned about in this paper. When $P=0$, that is to say, no AdS background, according to Eq.~(\ref{curvaturesp}), we can obtain the expression of the thermodynamic scalar curvature with respect to the temperature for Schwarzschild black hole $R_{_{\text{Schwarzschild}}}=-16\pi T_{_{\text{Schwarzschild}}}^2$. From the formula, we see that the thermodynamic scalar curvature is negative, which may be related to the information of attractive interaction between black hole molecules for the Schwarzschild black hole. On the other hand, based on the analysis of the thermodynamic geometric behavior of the RN black hole in our previous work \cite{Xu2019}, we have the expression of the thermodynamic scalar curvature with respect to the temperature for Schwarzschild black hole $R_{_{\text{Schwarzschild}}}=16\pi T_{_{\text{Schwarzschild}}}^2$. It is always positive which may be related to the information of repulsive interaction between black hole molecules for the Schwarzschild black hole. It's all about the Schwarzschild black hole, but we've got two opposite results, how do we understand them? According to the theory of black hole in general relativity, it is clear that the SAdS black hole is close to Schwarzschild black hole in small scale, while the RN black hole is similar to Schwarzschild black hole in large scale. To be clear, let's consider the behaviors of the temperature for Schwarzschild, SAdS, RN and RN-AdS black holes based on the expression of the temperature of RN-AdS black hole \cite{Kubiznak2012,Niu2012,Spallucci2013,Wang2019a}
\begin{eqnarray}\label{bht4}
T_{_{\text{RN-AdS}}}=\frac{1}{4\pi r}+2P r-\frac{Q}{4\pi r^3},
\end{eqnarray}
where $r$ is the horizon radius and $Q$ is square of charge of RN-AdS black hole. When $Q=0$, Eq.~(\ref{bht4}) is the temperature of SAdS black hole, and when $P=0$, it is that of the RN black hole, and when $P=0$ and $Q=0$, it degenerates to that of Schwarzschild black hole. In FIG. \ref{fig2}, we show the temperature curves of these four black holes. One can clearly see that the behavior of temperature of SAdS black hole is close to that of Schwarzschild black hole in small scale, while the the behavior of temperature of RN black hole is similar to that of Schwarzschild black hole in large scale. The SAdS black hole may be dominated by attractive interaction, while the RN black hole may be dominated by repulsive interaction \cite{Xu2019}. Therefore we can conclude that the Schwarzschild black hole may be dominated by repulsion on large scale and by attraction on small scale. In addition, we see that the temperature curves of the Schwarzschild black hole and the RN-AdS black hole have intersection. Hence we can approximate that this intersection is the transition point of repulsion and attraction for the Schwarzschild black hole.

Consequently we can approximately write an expression of the thermodynamic curvature of the Schwarzschild black hole as
\begin{eqnarray}\label{Schwarzschild}
R_{_{\text{Schwarzschild}}} &=16\pi T_{_{\text{Schwarzschild}}}^2 \text{sgn}\left(2P r-\dfrac{Q}{4\pi r^3}\right) \nonumber \\
& \nonumber \\
&=\left\{
             \begin{array}{ll}
             16\pi T_{_{\text{Schwarzschild}}}^2, & \text{large} ~r  \\
             & \\
             0, & 2P r=\dfrac{Q}{4\pi r^3}\\
             & \\
             -16\pi T_{_{\text{Schwarzschild}}}^2, & \text{small} ~r
             \end{array}
\right.
\end{eqnarray}

According to the behavior of the temperature of Schwarzschild black hole in FIG. \ref{fig2}, we can obtain that the Schwarzschild black hole may be dominated by repulsion on low temperature region and by attraction on high temperature region phenomenologically or qualitatively. However, some more precise and detailed microscopic analyses and how the thermodynamic curvature behavior of the Schwarzschild black hole will behave, especially on the intermediate temperature region (or intermediate scale), are still unknown. In addition, what principle drives the interaction between molecules of Schwarzschild black hole to present this pattern also needs to be further explored.

\section{Summary and Discussion}
Because the entropy and thermodynamic volume are not independent for spherically symmetric black holes in the AdS spacetime, it is belived that the most basic thermodynamic characteristic function is enthalpy $M(S,P)$, not internal energy $U(S,V)$. Based on this fact, we give the general form of the line element of Ruppeiner thermodynamic geometry of SAdS black hole and the specific forms in phase space $\{S,P\}$ and $\{T,V\}$. The results show that the thermodynamic curvatures obtained in the two phase spaces are equivalent, and the thermodynamic curvature is always negative which may be related to the information of attractive interaction between black hole molecules for the SAdS black hole. Meanwhile with the help of our results of SAdS black hole and the previous results of RN black hole, we also give an approximate expression of the thermodynamic curvature of the Schwarzschild black hole. This shows that the black hole may be dominated by repulsion on low temperature region and by attraction on high temperature region phenomenologically or qualitatively.

Here for the black holes with $C_{_V}=0$ in extended thermodynamics due to the non-independence of entropy and thermodynamic volume, we make some comments. We are very grateful to the anonymous referee for pointing out that the work \cite{Dolan2011b} first noted and resolved the problem of lack of independence of entropy and thermodynamic volume for non-rotating SAdS black hole. The author Dolan in \cite{Dolan2011b} dealt with rotating black holes, for which entropy and thermodynamic volume are independent and then take the limit of the angular momentum $J\rightarrow 0$. Like the SAdS black hole, RN-AdS black hole and GB-AdS black hole (and various other simple static black hole solutions of the pure Einstein gravity or higher-derivative generalizations thereof), when we deal with these black hole systems separately, we find that its thermodynamic volume is consistent with the expression of its geometric volume. Furthermore these black holes have zero $C_{_V}$ because the entropy and thermodynamic volume depend only on the horizon radius $r_h$, that is to say, a fixed thermodynamic volume $V$ means a fixed horizon radius $r_h$, resulting in a fixed entropy $S$. Hence in our current point of view, we consider the SAdS black hole as an independent thermodynamic system. We only need to deal with the system itself, and recognize that such a system has vanishing $C_{_V}$, without the aid of other black hole systems to analyze.

However the above phenomenon is no longer true for rotating black hole (for example Kerr-AdS black hole) and its thermodynamic volume does not look like any geometric volume \cite{Dolan2011b}. The same thing also happens with the STU-AdS \cite{Caceres2015,Johnson2019a}, Taub–NUT/Bolt-AdS \cite{Johnson2014} and charged (and exotic) BTZ black holes \cite{Johnson2019b,Cong2019}. When we adopt the correct procedure of calculating thermodynamic derivative first and then taking some parameter limits, like $J\rightarrow 0$ for rotating black holes in \cite{Dolan2011b}, although the form of solution can be returned to the case of Schwarzschild AdS black hole, we get the expression of thermodynamic volume which is completely different from the result of direct calculation from the Schwarzschild AdS black hole (which means this moment the thermodynamic volume and entropy are independent). We would like to thank anonymous referee for pointing out the key and interesting topic. This is probably due to the unclear definition of black hole volume, which is also a very significant research issue in black hole thermodynamics. For non-vanishing $C_{_V}$ systems (or the entropy and thermodynamic volume are independent), the first law is still $d M=T dS+V dP+\cdots$, where the ADM mass of black hole $M$ is the enthalpy. Our current thermodynamic geometric analysis based on enthalpy and Helmholtz free energy representations is still effective, but the difference is that the third term in Eq.~(\ref{linesp}) and the first term in Eq.~(\ref{linetv}) are no longer zero.

In addition, our present scheme can also be applied to other black holes, like RN-AdS and GB-AdS black holes \cite{Wei2019a,Wei2019b,Wei2019c,Niu2012,Wang2019a,Majhi2017,Majhi2019,Wang2019b}, to analyze some interesting behaviors of black holes in phase transition. Another topic of concern is the relationship between stability analysis of black holes and thermodynamic geometry. Because the entropy and thermodynamic volume of are not independent, the positive definiteness of the Hessian matrix may not be preserved. This issue is probably related to the thermal stability of black holes, which we are very concerned about and will be discussed in detail in the future.

\section*{Acknowledgments}
This research is supported by The Double First-class University Construction Project of Northwest University. The financial supports from National Natural Science Foundation of China (Grant No.11947208, Grant No.11605137, Grant No.11947301), Major Basic Research Program of Natural Science of Shaanxi Province (Grant No.2017KCT-12), Scientific Research Program Funded by Shaanxi Provincial Education Department (Program No.18JK0771) are gratefully acknowledged. The authors would like to thank the anonymous referee for the helpful comments that improve this work greatly.

\end{document}